\documentclass[seceq]{ptptex}

\usepackage{graphicx}
\newcommand{\lapprox}{\raisebox{-0.5ex}{$\
\stackrel{\textstyle<}{\textstyle\sim}\ $}}

\title{
Simulating Dense Matter
}

\author{
Simon \textsc{HANDS}
}

\inst{
Department of Physics, Swansea University, \\Singleton Park, 
Swansea SA2 8PP, U.K.
}

\abst{
I review the Sign Problem hindering lattice QCD simulations of dense baryonic
matter, focussing where possible on its physical relevance. The possibility 
of avoiding the Sign Problem via a duality transformation is also briefly
considered. Finally, I review evidence for deconfinement at
non-zero quark density in recent
simulations of Two Color QCD.
}

\begin{document}

\maketitle

\section{Motivation}

What is the nature of the QCD ground state in the limit $\mu_B/T\gg1$, where
$T$ is temperature and $\mu_B$ the baryon chemical potential? The
insight that diquark Cooper pair condensation in the color anti-triplet channel
is naturally promoted by one-gluon exchange suggests that in this 
r\'egime QCD is
a color superconductor. At asymptotic densities $\mu_B\to\infty$ where
weak-coupling methods can be trusted, the favoured ground state of 
QCD with three
light quark flavors~\footnote{Astrophysical arguments suggest that QCD
matter with four or more quark flavors cannot form a stable gravitationally
bound system.} exhibits Color-Flavor Locking, with 
spontaneous symmetry breaking pattern
\begin{equation}
SU(3)_c\otimes\mbox{SU(3)}_L\otimes\mbox{SU(3)}_R\otimes\mbox{U(1)}_B
\otimes U(1)_Q
\rightarrow\mbox{SU(3)}_\Delta\otimes U(1)_{\tilde Q}.
\end{equation}
Italicised and regular fonts denote respectively 
local and global symmetries of
massless quarks:
since both are spontaneously broken the CFL phase is simultaneously
superconducting and superfluid. 

At the densities available in stellar
cores, the QCD coupling $g(\mu_B)$ is no longer small, making reliable 
calculation difficult. In matter with $\mu_B\sim O(m_s)$ 
pairing may only take place between
$u$ and $d$ quarks, 
and further non-trivial constraints are imposed by requirements
of charge- and color-neutrality. Model approaches have predicted many exotic 
scenarios, such as gapless superconductivity, mixed states of normal and
superconducting matter, and crystalline LOFF phases~\cite{Igor}. 
The issue of which is the 
true ground state is ideally resolved, of course, by a systematic
non-perturbative lattice QCD calculation, as suggested by the talk's title.
It is worth recalling, however, that the most urgent question about quark matter
is whether it exists at all in our universe inside compact stars, or
whether the star would have collapsed into a black hole 
before the required core density
can be attained.  To settle this theoretically 
we need to solve the Tolman-Oppenheimer-Volkoff
equations for relativistic stellar structure, which requires quantitative
knowledge of the equation of state, ie quark density $n_q$, pressure $p$ and
energy density $\varepsilon$ as functions of $\mu_B$ for {\em all} 
$\mu_B>\mu_{Bo}$,
where $\mu_{Bo}\approx924$MeV is 
the onset value corresponding to self-bound nuclear matter. This issue, surely,
is the first goal of lattice QCD with $\mu_B\not=0$.

\section{The Sign Problem, and why we need it}
\label{sec:silverblaze}

Let me remind you why this problem was not solved years ago. In Euclidean 
metric the QCD Lagrangian is written
\begin{equation}
{\cal L}_{QCD}=\bar\psi M\psi+{1\over4}F_{\mu\nu}F_{\mu\nu}\;\;\;
\mbox{with}\;\;\;M(\mu)=D{\!\!\!\! /\,}[A]+\mu\gamma_0+m,
\end{equation}
where $\mu=\mu_B/N_c$ is the quark chemical potential. It is straightforward
to show $\gamma_5M(\mu)\gamma_5\equiv M^\dagger(-\mu)$ implying
$\mbox{det}M(\mu)=(\mbox{det}M(-\mu))^*$, so that the path integral measure is
not positive definite for $\mu\not=0$. This is not solely an issue for fermions;
it can be traced to the explict breaking of time reversal
symmetry by the term with $\mu\not=0$, which in Euclidean metric corresponds to
a breaking of the symmetry under $i\mapsto-i$. The consequences are drastic;
Monte Carlo importance sampling, the mainstay of lattice QCD, becomes
ineffective.

To see why consider the formal solution to the Sign Problem known as
{\em reweighting}. Here the phase of $\mbox{det}M$ is treated as an observable,
and expectation values defined by
\begin{equation}
\langle{\cal O}\rangle={{\langle\langle{\cal
O}\mbox{arg}(\mbox{det}M)\rangle\rangle}\over
{\langle\langle\mbox{arg}(\mbox{det}M)\rangle\rangle}},
\label{eg:reweight}
\end{equation}
with $\langle\langle\ldots\rangle\rangle$ defined using the positive measure
$\vert\mbox{det}M\vert e^{-S_{boson}}$. Unfortunately, both numerator and
denominator of (\ref{eg:reweight}) 
are exponentially suppressed as $V\to\infty$, eg:
\begin{equation}
\langle\langle\mbox{arg}(\mbox{det}M)\rangle\rangle={{Z_{true}}\over{Z_{fake}}}
=\exp(-\Delta F)\sim\exp(-\#V)
\end{equation}
where in the last step we assume the free energy $F$ is extensive. On general
grounds we expect any signal for $\langle{\cal O}\rangle$ to be overwhelmed by 
statistical noise in the thermodynamic limit.

\begin{figure}[tbh]
\centering
\includegraphics[width=5.0cm]{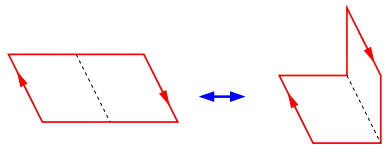}
\caption{Two polymers of opposite sign.}\label{fig:flip}
\end{figure}
It is instructive and surprisingly easy to introduce a Sign Problem into 
QCD at $\mu=0$. Consider the {\em polymer representation}~\cite{KST} for the 
QCD partition function:
\begin{eqnarray}
{\cal Z}_{QCD}&=&\int\!\! DU\,\mbox{det}M[U;m]e^{-S_W[U]}\nonumber\\
&\propto&\int\!\! DU\,\sum_{\{{\cal C}\}}(2m)^{{N_m}}(-1)^{N_\Gamma}
\prod_{\Gamma\in{\cal C}}\Bigl((\mbox{tr}\prod_{\ell\in\Gamma}\gamma_\ell)
(\mbox{tr}\,U_\Gamma)\Bigr) e^{-S_W[U]}.
\end{eqnarray}
Each non-vanishing term in the expansion of the determinant is
represented by a partition ${\cal C}$
of the lattice into $N_m$ monomers, $N_d$ dimers and
$N_\Gamma$ polymers, the latter defined as oriented closed paths of links. In
the strong-coupling limit where only monomers and dimers contribute each term is
positive.
Polymers
contribute not only a Wilson loop $U_\Gamma$ to the effective action but also a
shape-dependent sign factor $-\mbox{tr}\prod_{\ell\in\Gamma}\gamma_\ell$,
where $\gamma_\ell=\pm\gamma_\mu$ depending on whether the link $\ell$ points
along $\pm\hat\mu$. The resulting Sign Problem makes simulation of even the
non-interacting system difficult~\cite{Istvan}.
At weak gauge coupling it is tempting to interpret the
polymers as quark worldlines, but note that the overall sign contributed by
${\cal C}$ can be changed by the innocent-looking flip shown in 
Fig.~\ref{fig:flip}. It is difficult to believe that in this limit there can be
any significant correlation between the sign of ${\cal C}$ and 
long-range physics.


The positive measure $\mbox{det}M^\dagger M$ used in practical fermion
algorithms describes color triplet quarks $q$ and color anti-triplet conjugate
quarks $q^c$. There are thus gauge-invariant $qq^c$ bound states 
with baryon number $B\not=0$. At $\mu=0$ we are content to consider these states
as extra ``mesons'' and move on. Once $\mu\not=0$, however, this position is
untenable. The lightest baryon in this model's spectrum is degenerate with the
pion, so that there is an unphysical onset transition between vacuum and
baryonic matter at $\mu_o\simeq{1\over2}m_\pi$. Only calculations performed
with the correct complex measure $\mbox{det}^2M$ can yield cancellations among 
configurations with differing phases, which nullify the effect of $qq^c$ states
and postpone the onset transition to the phenomenologically-observed
$\mu_o\approx{1\over3}m_N$. For the vacuum to persist as the correct ground
state in the range 
$\mu\in({1\over2}m_\pi,{1\over3}m_N)$
it seems we actually need a Sign Problem.

This cancellation has been numerically verified in simulations of Two Color QCD
(QC$_2$D)
with a single staggered quark flavor in the adjoint representation, where it was
found that the signal for a fake transition at $\mu\simeq{1\over2}m_\pi$,
to a superfluid phase whose order parameter $\langle qq\rangle$ vanishes
identically due to the Pauli principle,
went away once the sign of the determinant was correctly taken into
account~\cite{HMSS}. More recently a visualisation of the Sign Problem has
emerged from an analytic solution of a random matrix model with the same global
symmetries as QCD, corresponding to the so-called mesoscopic limit of
$V\to\infty$ with $m_\pi^2f_\pi^2V$ fixed~\cite{RMT}. The chiral condensate 
can be expressed in terms of the distribution $\rho(z)$ of eigenvalues of 
$M-m$ in the complex plane:
\begin{equation}
\langle\bar\psi\psi\rangle=\lim_{m\to0}\lim_{V\to\infty}V^{-1}\int dzd\bar z
{{\rho(z,m;\mu)}\over{z+m}}.
\label{eq:BC}
\end{equation}
For $\mu=0$ or the quenched limit $N_f=0$, $\rho$ is real, but in
general it is a complex-valued function. The explicit solution for $\mu>m_\pi/2$
shows that for $\mbox{Re}\;z>m$, $\rho$ develops an oscillatory structure, with 
a characteristic wavelength of $O(V^{-1})$ and amplitude $O(e^V)$. Any function
calculated using a formula such as (\ref{eq:BC}) must receive wildly fluctuating
contributions from different regions of the plane, but remarkably, it can be
shown that the result behaves entirely in accord with physical expectations,
namely that $\langle\bar\psi\psi(m)\rangle$ changes sign at $m=0$, 
but exhibits no sign of discontinuous behaviour as $\mu$ passes though
$m_\pi/2$~\cite{RMT2}.

\section{High Density Effective Theory and an Optimistic Conjecture}

\begin{figure}[tbh]
\centering
\includegraphics[width=5.0cm]{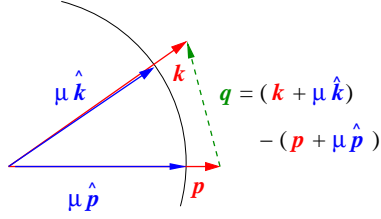}
\caption{Kinematics of quasiquark-gluon scattering at the Fermi surface.}
\label{fig:scatter}
\end{figure}
For $\mu\gg T$, $\mu\gg\Lambda_{QCD}$, QCD is supposed to exist in a deconfined
phase of degenerate weakly interacting quarks.
Is it possible to construct an effective theory in terms of quasiparticle
degrees of freedom at the Fermi surface, with momenta $p,k\ll\mu$?
Define ``fast'' (--) and ``slow'' (+) degrees of freedom via the 
decomposition:
\begin{equation}
\psi(x)=\exp(i\mu{\bf x}.\hat{\bf p})[\psi_+(x)+\psi_-(x)]\;\;
\mbox{with}\;\;\psi_\pm(p)={1\over2}(1\pm\vec\alpha.\hat{\bf p})\psi(p).
\end{equation}
The phase and projection factors ensure that the $\psi_\pm$ fields scatter off 
gluons with physical momenta $q$, according to the kinematics shown in
Fig.~\ref{fig:scatter}. The quark Lagrangian in Minkowski metric now reads
\cite{DKH}
\begin{equation}
\bar\psi_+\gamma_0(1,\hat{\bf p})^\nu i\tilde D_\nu\psi_+
+\bar\psi_-\gamma_0[(1,-\hat{\bf p})^\nu i\tilde D_\nu+2\mu]\psi_-
+[\bar\psi_-i\gamma_\perp^\nu\tilde D_\nu\psi_++h.c.],
\end{equation}
where $\tilde A_\nu=e^{-i\mu{\bf x.\hat p}}A_\nu e^{i\mu{\bf x.\hat p}}$.
At tree level the massive $\psi_-$ field can be integrated out to yield
\begin{equation}
{\cal L}_{HDET}=\bar\psi_+i\gamma_\parallel^\nu\tilde D_\nu\psi_+
+{g^2\over{2\mu}}\bar\psi_+\gamma_0(\gamma_\perp^\nu\tilde A_\nu)^2\psi_+
+O(D^3\mu^{-2})
\end{equation}
with $\gamma_\parallel^\nu=(\gamma_0,\hat{\bf p}\vec\gamma.\hat{\bf p})^\nu$,
$\gamma_\perp^\nu=\gamma^\nu-\gamma_\parallel^\nu$. 

The resulting theory has
been used with some success to analyse the color superconducting phase; however,
for our purposes it is more interesting to continue to Euclidean
metric~\cite{HH}. Since $\gamma_{\parallel\nu}\tilde D_\nu$ is anti-hermitian
and satisfies $\{\gamma_{\parallel\nu}\tilde D_\nu,\gamma_5\}=0$, it is
straightforward to show $\mbox{det}(\gamma_{\parallel\nu}\tilde D_\nu)$ is
positive definite. There is therefore
no Sign Problem in the limit $\mu\to\infty$!

I have argued the Sign Problem is intractable almost everywhere in the $(\mu,T)$
plane as $V\to\infty$~\footnote{An exception appears to be exactly at $T=0$
below the fake onset at $\mu=m_\pi/2$~\cite{SV}.}. 
However, it is perhaps possible to
distinguish between regions such as $\mu\in({1\over2}m_\pi,{1\over3}m_N)$ where
we know from Sec.~\ref{sec:silverblaze} that sign cancellations are both subtle
and crucial to obtaining physically sensible predictions, from regions where the
sign fluctuations are not so strongly correlated with long range physics. One
such region appears to be the upper left-hand corner of the QCD phase diagram
$\mu/T\lapprox1$ where RHIC physics takes place: on finite volumes
there is a pleasing
consistency between approaches based on reweighting
(which must inevitably fail in
the thermodynamic limit) and alternative methods based on analytic
continuation~\cite{LP}. While a systematic numerical treatment of HDET is yet 
to emerge, the previous paragraph at least suggests that in the cold dense 
r\'egime
$\mu/T\gg1$ a ``solution'' of the Sign Problem may not be a crucial component of
the physics, and that it may be possible to perform controlled calculations on
reasonable volumes. Another approach, which I will pursue in
Sec.~\ref{sec:TCQCD} below, is to
argue that a theory with no Sign Problem such as
QC$_2$D still models much relevant physics in this r\'egime.

\section{Are we using the Right Basis?}

Large cancellations between either Feynman diagrams or gauge
configurations hint at low
calculational efficiency. Maybe gauge covariant quarks and gluons are not the
best degrees of freedom at high density? It is possible in
some cases to effect a transformation to another basis in which the Sign Problem
is absent. An intriguing example
illustrating this comes from 3$d$ scalar QED~\cite{SQED}, with Lagrangian
\begin{equation}
{\cal L}_{SQED}={1\over4}F^2+\vert
D\phi\vert^2+m^2\vert\phi\vert^2+\lambda\vert\phi\vert^4-
{1\over2}\varepsilon_{ijk}H_iF_{jk},
\label{eq:SQED}
\end{equation}
where $H$ is a real vector source coupled to a generalised
$B$-field. The model has a
transition separating Coulomb and Higgs phases which is second order for
sufficiently large $\lambda/e^2$. There is a conjectured duality at this
critical point with a complex scalar field theory described by
\begin{equation}
{\cal L}_{SFT}=[(\partial-\tilde eH)_k\tilde\phi^*][(\partial+\tilde
eH)_k\tilde\phi]+\tilde
m^2\vert\tilde\phi\vert^2+\tilde\lambda\vert\tilde\phi\vert^4+\cdots
\label{eq:SFT}
\end{equation}
The point is that $\tilde eH_3$ with $\tilde e=2\pi/e$ is a real chemical
potential associated with the conserved charge density
$2\mbox{Im}(\tilde\phi^*\partial_3\tilde\phi)$. The Lagrangian (\ref{eq:SFT})
is in general complex, describing planar Bose-Einstein condensation (BEC) of
$\tilde\phi$-quanta for 
$\tilde eH_3\approx\tilde m$. In principle, however, it could be studied via 
simulations of the real action (\ref{eq:SQED}).

A more recent example due to Endres~\cite{Endres} exploits an exact duality
between complex scalar field theory in $d$ dimensions and a loop gas:
\begin{eqnarray}
{\cal Z}(\mu)
&=&\int\!\!D\phi
D\phi^*\exp\Bigl[-\sum_{x\nu}(2\phi^*_x\phi^{\phantom *}_x
-\phi^*_xe^{\mu\delta_{\nu,0}}\phi^{\phantom *}_{x+\hat\nu}
-\phi^*_{x+\hat\nu}e^{-\mu\delta_{\nu,0}}\phi^{\phantom *}_x)\nonumber\\
&&\phantom{=\int\!\!D\phi
D\phi^*\exp\Bigl[-\sum_{x\nu}(2\phi^*_x\phi^{\phantom *}_x
-\phi^*_xe^{\mu\delta_{\nu,0}}\phi^{\phantom *}_{x+\hat\nu}
}
-m^2\sum_x\phi^*_x\phi^{\phantom *}_x
-\sum_x V(\phi^*_x\phi^{\phantom *}_x)\Bigr]\nonumber\\
&\propto&\sum_{\{\ell\}}\int_0^\infty \!\!\!\!\rho D\rho\,
e^{\mu\sum_x\ell_{x,0}}
\prod_x\Bigl[e^{-(2d+2+m^2)\rho_x^2-V(\rho_x^2)}\prod_\nu
I_{\ell_{x,\nu}}(2\rho_x\rho_{x+\hat\nu})\Bigr],
\label{eq:loopgas}
\end{eqnarray}
where $\rho_x=\vert\phi_x\vert$, and we have introduced integer-valued 
link variables $\ell_{x,\nu}$ governed by the constraint
$\partial^-_\nu\ell_\nu=0$. Once again ${\cal Z}$ is recast in terms of a
functional integral over real variables, and the Sign Problem averted.
Remarkably, it has proved possible to simulate the action (\ref{eq:loopgas})
efficiently, yielding non-trivial results of physical interest~\cite{Endres}.

\section{Two Color Matters}
\label{sec:TCQCD}

QCD with gauge group SU(2) and an even number of 
fundamental quark flavors has a real functional measure even once $\mu\not=0$,
and remains the {\em only} dense matter system with long-range fundamental
interactions amenable to study with orthodox lattice methods. Since $q$ and
$\bar q$ live in equivalent representations of the color group, hadron
multiplets contain both $q\bar q$ mesons and $qq$, $\bar q\bar q$ baryons. Near
the chiral limit at $\mu=0$ we expect spontaneous  chiral symmetry breaking
implying $m_\pi\ll m_\rho$, where $\rho$ denotes the next lightest hadron. The
theory's $\mu$-dependence for $\mu<m_\rho$ 
can be analysed using chiral perturbation theory~\cite{KSTVZ} ($\chi$PT).
The key result is that a second order onset transition occurs at
$\mu_o={1\over2}m_\pi$ to a phase with quark charge density $n_q>0$. 
For $N_f=2$ the matter
which forms is composed of tightly-bound scalar diquarks, which Bose condense
to form a gauge-invariant superfluid BEC $\langle qq\rangle\equiv
\langle\psi^{tr}C\gamma_5\sigma_2\tau_2\psi\rangle\not=0$, where the matrices
act on spinor, flavor and color indices respectively. Since $n_q\to0$ as
$\mu\searrow\mu_{o}$ the matter can be arbitrarily dilute.

The $\chi$PT predictions for $\langle\bar\psi\psi(\mu)\rangle$ and $\langle
qq(\mu)\rangle$ have been confirmed by lattice simulations with staggered
fermions~\cite{TCQCD,HMSS}. For our purposes the most relevant prediction
is the equation of state for $T=0$, $\mu>\mu_o$:
\begin{equation}
n_{\chi PT}=8N_ff_\pi^2\mu\left(1-{\mu_o^4\over\mu^4}\right),
\end{equation}
leading to the pressure $p_{\chi PT}=\int_{\mu_o}^\mu n_qd\mu$ 
and energy density 
$\varepsilon_{\chi PT}=-p+\mu n_q$:
\begin{equation}
p_{\chi PT}
=4N_ff_\pi^2\left(\mu^2+{\mu_o^4\over\mu^2}-2\mu_o^2\right);\;\;
\varepsilon_{\chi PT}=
4N_ff_\pi^2\left(\mu^2-3{\mu_o^4\over\mu^2}+2\mu_o^2\right).
\end{equation}
This is to be contrasted with another paradigm for cold dense matter, namely 
a degenerate system of weakly interacting quarks populating a Fermi sphere up
to some maximum momentum $k_F\approx E_F=\mu$, obeying Stefan-Boltzmann
(SB) scaling:
\begin{equation}
n_{SB}={{N_fN_c}\over{3\pi^2}}\mu^3;\;\;\;\varepsilon_{SB}=3p_{SB}=
{{N_fN_c}\over{4\pi^2}}\mu^4.
\label{eq:SB}
\end{equation}
The appearance of $N_c$ underlines that (\ref{eq:SB}) describes a deconfined
phase. Superfluidity in this scenario arises from condensation of diquark 
Cooper pairs within a layer of thickness $\Delta$ centred on the Fermi surface,
implying $\langle qq\rangle\propto\Delta\mu^2$.
\begin{figure}[htb]
\begin{minipage}[t]{70mm}
\begin{center}
\includegraphics[width=5.5cm]{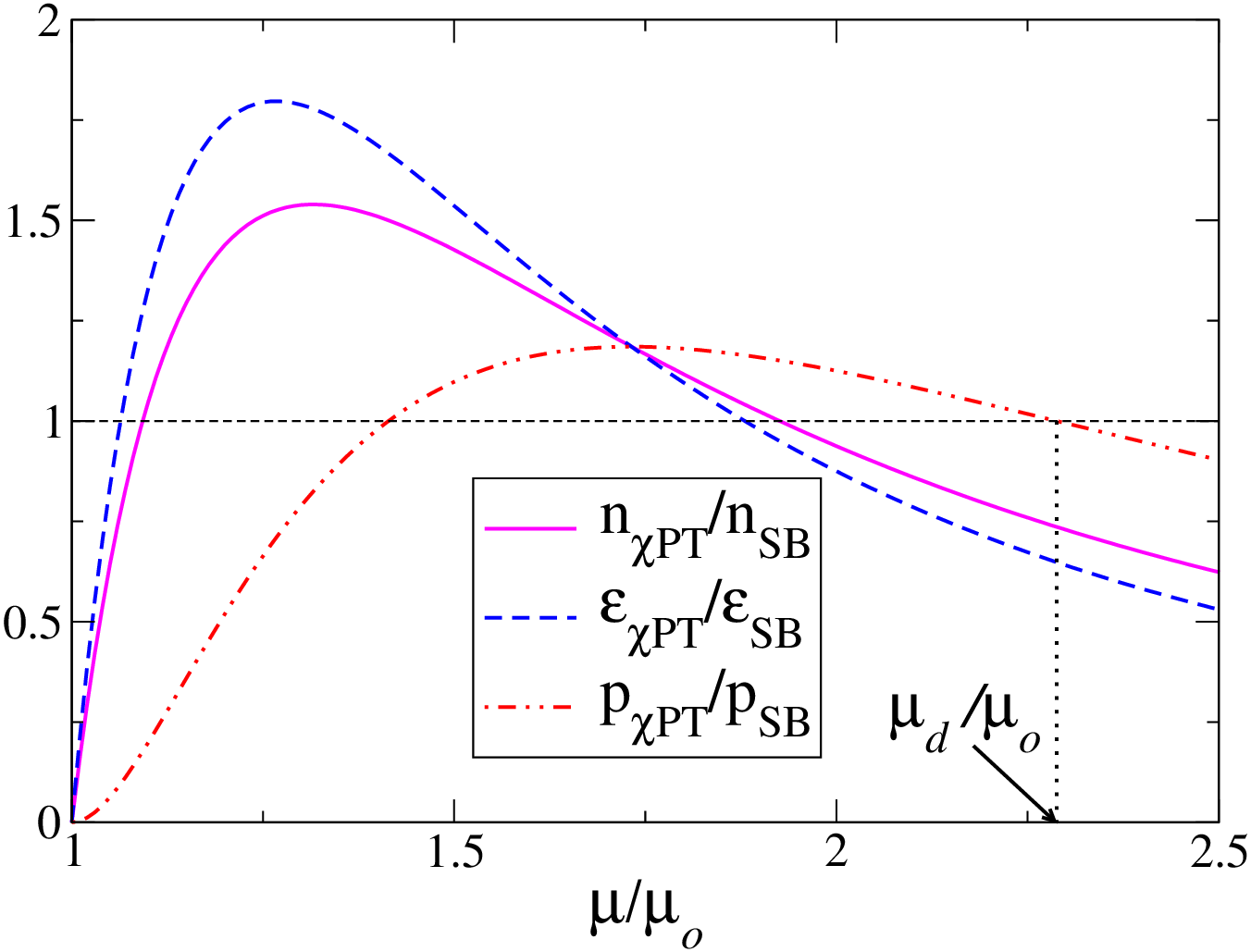}
\caption{Comparison of $\chi$PT with free quarks.}
\label{fig:model}
\end{center}
\end{minipage}
\hspace{\fill}
\begin{minipage}[t]{70mm}
\begin{center}
\includegraphics[width=5.5cm]{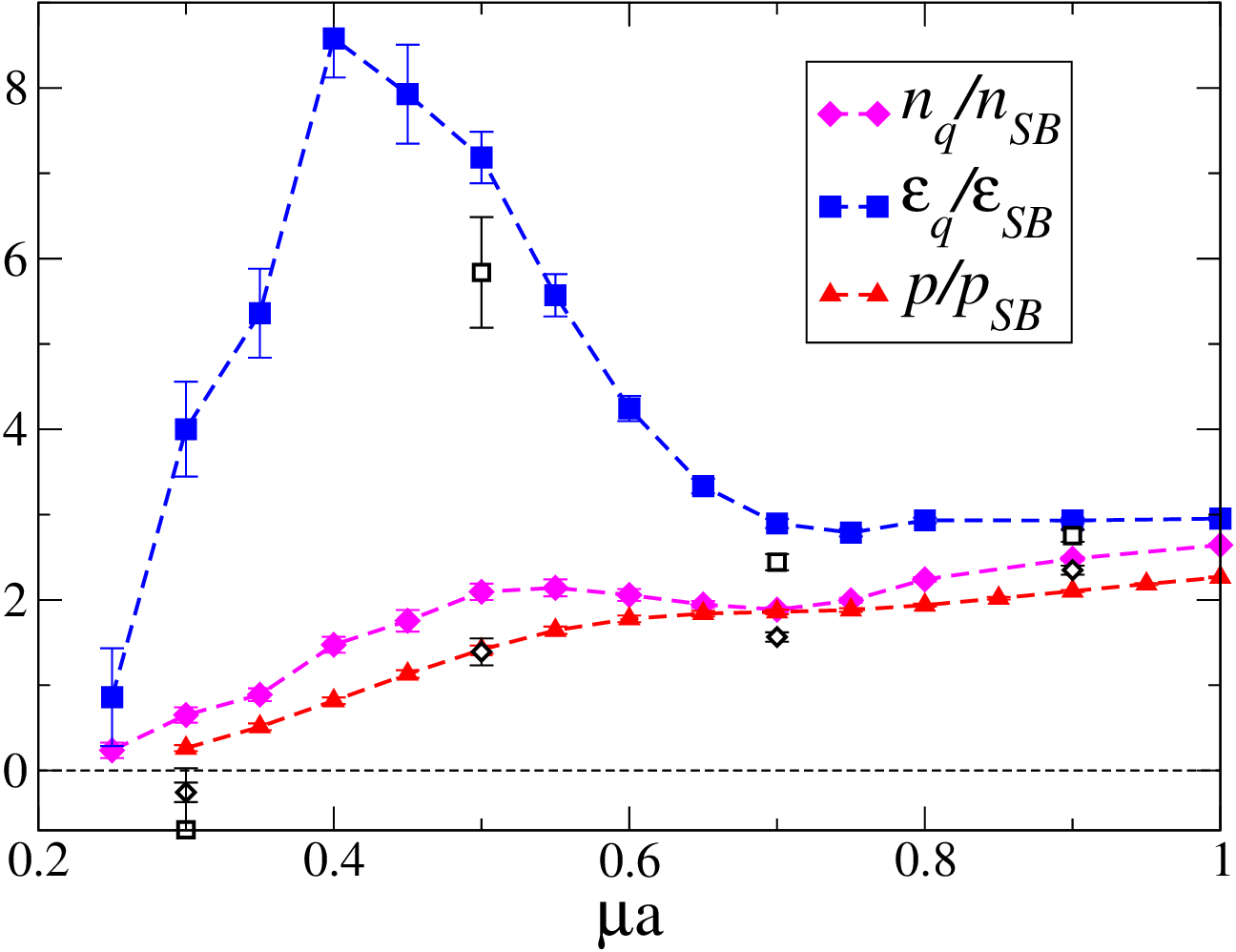}
\caption{Equation of state from lattice QC$_2$D.}
\label{fig:eneps}
\end{center}
\end{minipage}
\end{figure}
Fig.~\ref{fig:model} shows the ratio of $\chi$PT to SB predictions as functions
of $\mu/\mu_o$ for the choice $f_\pi^2=N_c/6\pi^2$. By equating pressures, we
naively predict a first-order deconfining transition from BEC to quark matter
at $\mu_d/\mu_o\approx2.3$~\footnote{The apparent transition at
$\mu_d/\mu_o\approx1.4$ can be eliminated by introducing a bag constant to
stabilise the confined phase at low density.}.

To test whether this prediction is robust we have performed lattice 
simulations of QC$_2$D using $N_f=2$ Wilson
quarks~\cite{HKS}. 
Initial runs have been on a $8^3\times16$ lattice with lattice
spacing (defined via the string tension) $a\simeq0.22$fm, corresponding to 
$T\approx60$MeV. We have not attempted to get particularly close to the chiral
limit; $m_\pi a=0.79(1)$ and $m_\pi/m_\rho=0.80(1)$. The code's only novelty 
is the inclusion of a diquark source term of the form $j(qq+\bar
q\bar q)$, with $ja=0.04$ for the most part, 
which both ensures ergodicity and regularises IR fluctuations in the
superfluid phase. Fig.~\ref{fig:eneps} shows the resulting curves for
$n_q$, $p$ and $\varepsilon_q$ in the same format as Fig.~\ref{fig:model} (open
symbols denote the $j\to0$ extrapolation). 
There appears to be a transition from
confined  bosonic ``nuclear matter'' to deconfined quark matter at
$\mu_da\approx0.65$. For large $\mu$, $n_q/n_{SB}\approx p/p_{SB}\approx2$,
consistent with a bound system having $E_F<k_F$.
\begin{figure}[tbh]
\centering
\includegraphics[width=5.5cm]{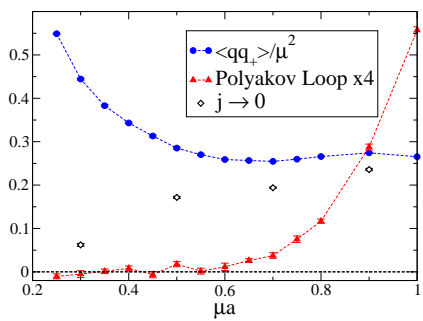}
\caption{Evidence for deconfinement.}
\label{fig:orderps}
\end{figure}
The claim is supported by Fig.~\ref{fig:orderps}, which shows the Polyakov
line rise from zero at $\mu_d$, coincident with the superfluid order
parameter assuming the scaling $\langle qq\rangle\propto\mu^2$ consistent with
Cooper pairing at a Fermi surface. A condensed matter physicist would refer to
this as a BEC/BCS crossover. Similar conclusions have been reached in a study
of topological charge susceptibility using staggered fermions with
$N_f=8$~\cite{AEL}.

\begin{figure}[htb]
\begin{minipage}[t]{70mm}
\begin{center}
\includegraphics[width=5.5cm]{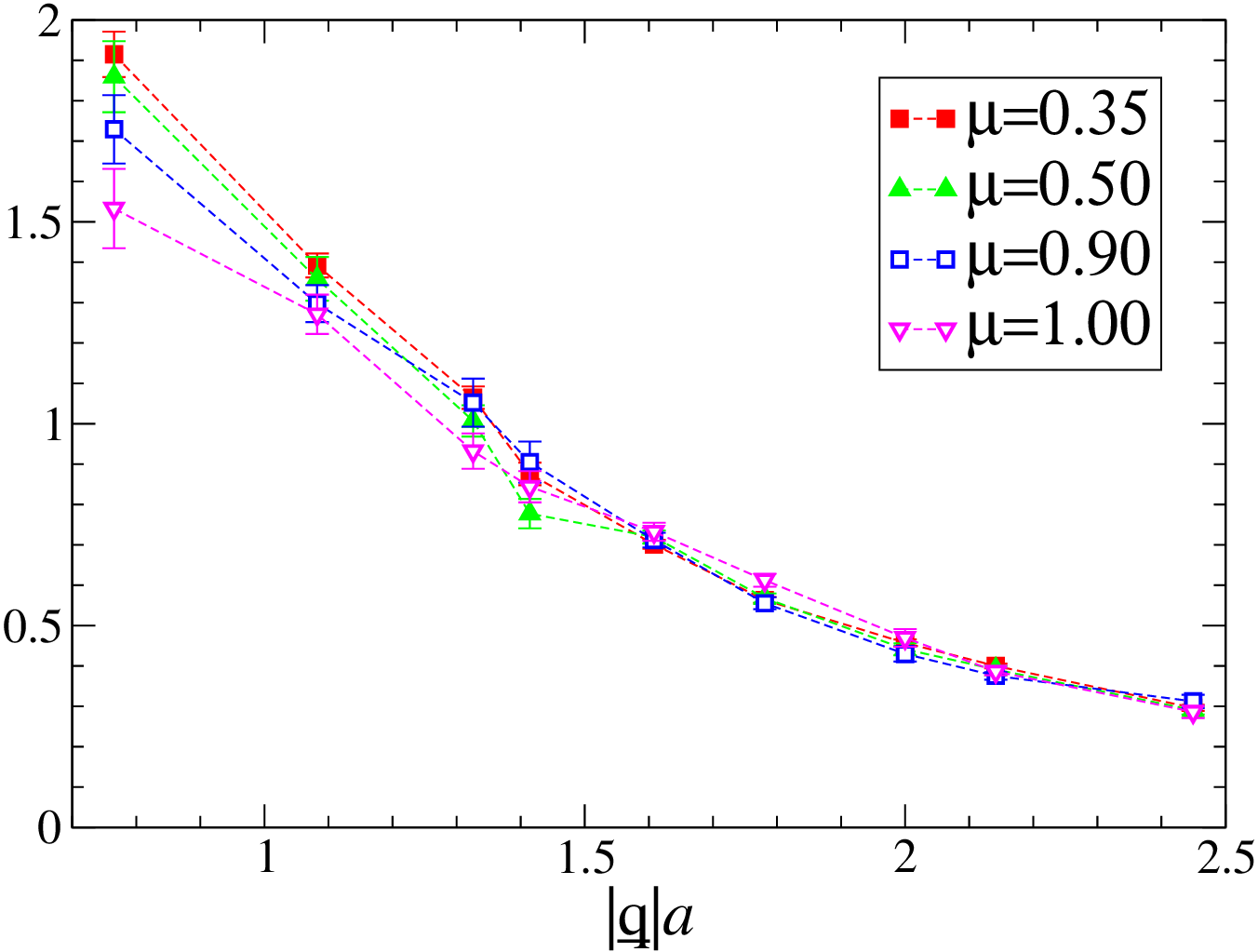}
\caption{Electric gluon $D_E$ in the static limit.}
\label{fig:DE_k0}
\end{center}
\end{minipage}
\hspace{\fill}
\begin{minipage}[t]{70mm}
\begin{center}
\includegraphics[width=5.5cm]{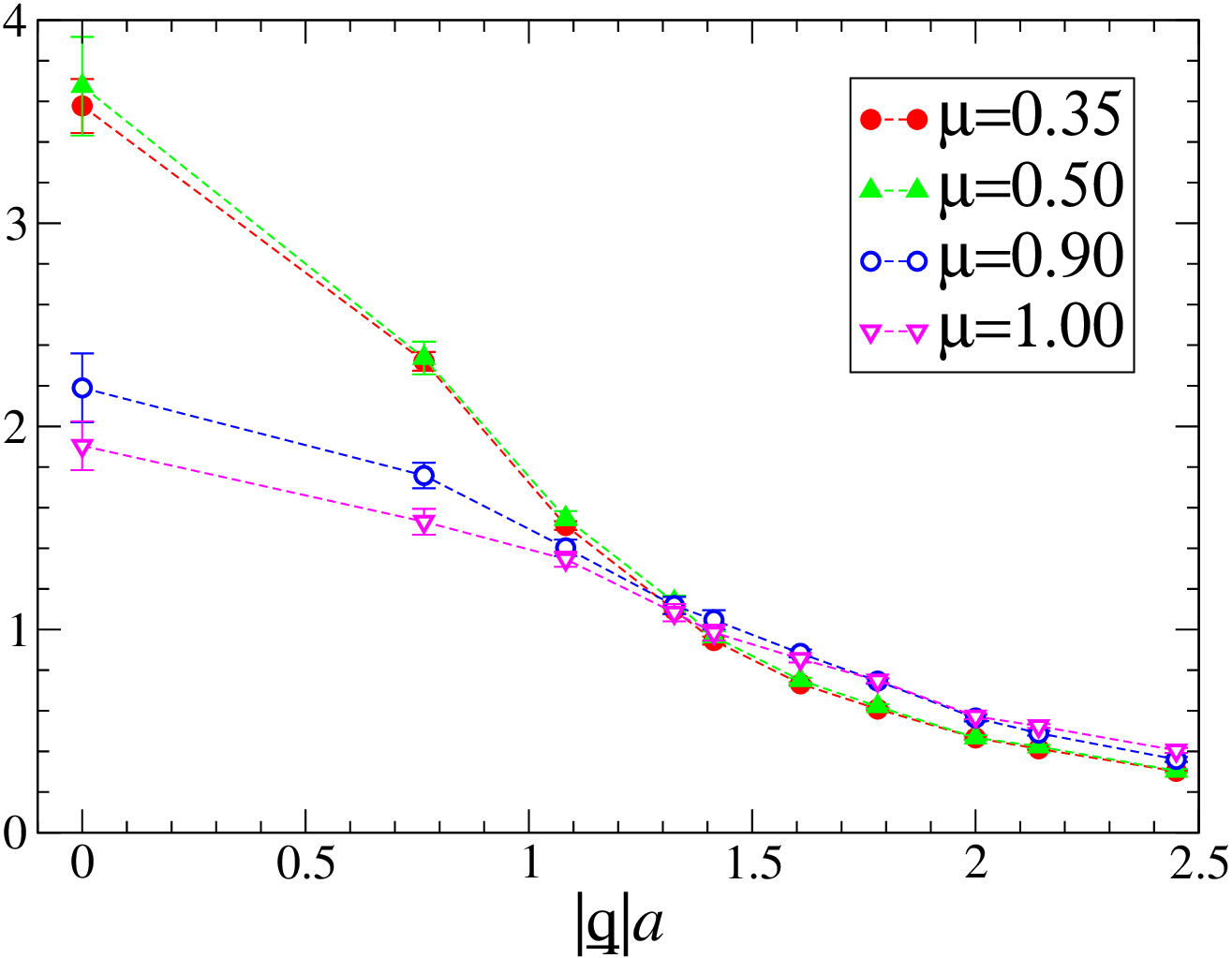}
\caption{Magnetic gluon $D_M$ in the static limit.}
\label{fig:DM_k0}
\end{center}
\end{minipage}
\end{figure}
A major motivation for studying QC$_2$D is to understand
how deconfined quarks affect gluodynamics; as argued above, it is at least
plausible that the lessons learned may apply to physical QCD. In any medium 
with a preferred rest frame, the gluon propagator can be decomposed 
as follows:
\begin{equation}
D_{\mu\nu}(q)={\cal P}^M_{\mu\nu}D_M(q_0^2,\vec q.\vec q)+
{\cal P}^E_{\mu\nu}D_E(q_0^2,\vec q.\vec q)+
{\cal P}^L_{\mu\nu}D_L(q_0^2,\vec q.\vec q)
\end{equation}
with
\begin{equation}
{\cal P}_{ij}^M=\delta_{ij}-{{q_iq_j}\over{\vec q.\vec q}},\;\;
{\cal P}_{00}^M={\cal P}_{0i}^M=0;\;\;\;
{\cal P}_{\mu\nu}^E=\delta_{\mu\nu}-{{q_\mu q_\nu}\over{q_0^2+\vec q.\vec q}}
-{\cal P}_{\mu\nu}^M;\;\;\;
{\cal P}_{\mu\nu}^L={{q_\mu q_\nu}\over{q_0^2+\vec q.\vec q}}.
\nonumber
\end{equation}
We have used Landau gauge in which $D_L=0$. Figs.~\ref{fig:DE_k0} and
\ref{fig:DM_k0} plot $D_E$ and $D_M$ in the static limit $q_0=0$ as
functions of $\vert\vec q\vert$, for various $\mu$ on either side of the 
deconfinement transition. In the electric sector for
$\mu a\geq0.9$ Fig.~\ref{fig:DE_k0}
shows evidence for some Debye screening as $\vert\vec q\vert\to0$.
Deconfinement has a much more dramatic effect in the magnetic sector shown in
Fig.~\ref{fig:DM_k0}, where in the same limit
the propagator is screened by $O$(50\%). This is
significant because in perturbation theory magnetic gluons are not
screened in the static limit; indeed, this is at the origin of the celebrated
scaling $\Delta\propto
e^{-\#/g}$ of the color superconducting gap predicted by weak-coupling
methods~\cite{Son}.

In summary, even models with no Sign Problem may hold surprises for us
at large $\mu$. It would, of course, be nice to compare lattice results for 
QC$_2$D with other non-perturbative approaches such as Schwinger-Dyson
equations. My overall feeling, though, is that a radical reformulation of 
non-perturbative QCD is needed before numerical approaches can make further
headway.

\section*{Acknowledgements}
It is a pleasure to thank my hosts for organising such an excellent meeting.
My trip to YKIS2006 was supported by the Royal Society.


\end{document}